\def\Phij{{\widetilde{\Phi}_j}}
\def\Phii{{\widetilde{\Phi}_i}}
\def\Phil{{\widetilde{\Phi}_1}}
\def\Phir{{\widetilde{\Phi}_2}}
\def\Tj{{\widetilde{T}_j}}
\def\Tl{{\widetilde{T}_1}}
\def\Tr{{\widetilde{T}_2}}
\def\Inorm{{\widetilde{I}_c}}

\documentclass[aps,prappl,twocolumn,groupedaddress,superscriptaddress,floatfix,longbibliography]{revtex4-1}
\usepackage{epsfig}
\usepackage{multirow}
\usepackage{amsmath,amssymb,mathtools}
\usepackage{color}
\usepackage{graphicx}
\usepackage{dcolumn}
\usepackage{bm}
\usepackage{subfigure}
\usepackage[utf8]{inputenc}
\usepackage[T1]{fontenc}
\usepackage{tikz}
\usepackage{upgreek}
\usepackage{amsmath}

\begin{document}

\title{Temperature-biased double-loop Josephson flux transducer}

\author{C. Guarcello}
\affiliation{Dipartimento di Fisica ``E.R. Caianiello'', Universit\`a di Salerno, Via Giovanni Paolo II, 132, I-84084 Fisciano (SA), Italy}
\affiliation{INFN, Sezione di Napoli Gruppo Collegato di Salerno, Complesso Universitario di Monte S. Angelo, I-80126 Napoli, Italy}
\author{R. Citro}
\affiliation{Dipartimento di Fisica ``E.R. Caianiello'', Universit\`a di Salerno, Via Giovanni Paolo II, 132, I-84084 Fisciano (SA), Italy}
\affiliation{INFN, Sezione di Napoli Gruppo Collegato di Salerno, Complesso Universitario di Monte S. Angelo, I-80126 Napoli, Italy}
\affiliation{CNR-SPIN c/o Universit\'a degli Studi di Salerno, I-84084 Fisciano (Sa), Italy}
\author{F. Giazotto}
\affiliation{NEST, Istituto Nanoscienze-CNR and Scuola Normale Superiore, Piazza San Silvestro 12, I-56127 Pisa, Italy}
\author{A. Braggio}
\affiliation{NEST, Istituto Nanoscienze-CNR and Scuola Normale Superiore, Piazza San Silvestro 12, I-56127 Pisa, Italy}

\date{\today}

\begin{abstract}
We theoretically study the behavior of the critical current of a thermally-biased tunnel Josephson junction with a particular design, in which the electrodes of the junction are enclosed in two different superconducting loops pierced by independent magnetic fluxes. In this setup, the superconducting gaps can be modified independently through the magnetic fluxes threading the loops. We investigate the response of the device as a function of the magnetic fluxes, by changing the asymmetry parameter, i.e., the ratio between the zero-temperature superconducting gaps $\delta=\Delta_{10}/\Delta_{20}$, and the temperatures of the two rings. We show a magnetically controllable step-like response of the critical current, which emerges even in a symmetric junction, $\delta=1$. Finally, we discuss the optimal working conditions and the high response of the critical current to small changes in the magnetic flux, reporting good performances of a magnetic flux-to-critical current transducer, with a high transfer function that depends on the operating point, the temperature gradient, and the quality of the junction.
\end{abstract}

\maketitle

\section{Introduction}
\label{section0}\vskip-0.2cm

Recently, thermal transport at the nanoscale
and the research field of superconducting phase coherent caloritronics are attracting interest~\cite{Mes06,Muh12,Sot13,Gas15,ForGia17,Pek19,Hwa20,Mai20,Dai21,Pek21} due to the importance for the whole field of quantum technologies and quantum computing. In this context, it was recently argued that the Josephson transport~\cite{GuaBra19,Gua19,MarPRR20,Ave21} across a superconductor-insulator-superconductor (SIS) Josephson junction (JJ), formed by different superconductors residing at different temperatures, can show unexpected, and quite peculiar, features. For example, strong nonlinear temperature biases can cause a spontaneous breaking of the particle-hole symmetry, generating even a bipolar thermoelectric effect, when Josephson coupling is suppressed~\cite{MarPRL20,MarPRB20}. Intriguing phenomena have been predicted also in the Josephson transport, such as a sudden variation of the critical current in response to a temperature bias~\cite{GuaBra19}. Surprisingly, the critical current can even increase with the temperature, thus suggesting to apply this phenomenon for a wide-band threshold calorimeter~\cite{Gua19}.

All the cases described so far underline the importance of the alignment mechanism of the two superconducting singularities in the density of state (DOS) and its anomalous components. This matching of the singularities can be triggered by different mechanisms determined by temperatures~\cite{GuaBra19}, biasing~\cite{MarPRL20}, and/or exchange fields~\cite{Ger21}. 

\begin{figure}[b!!]
\centering
\includegraphics[width=\columnwidth]{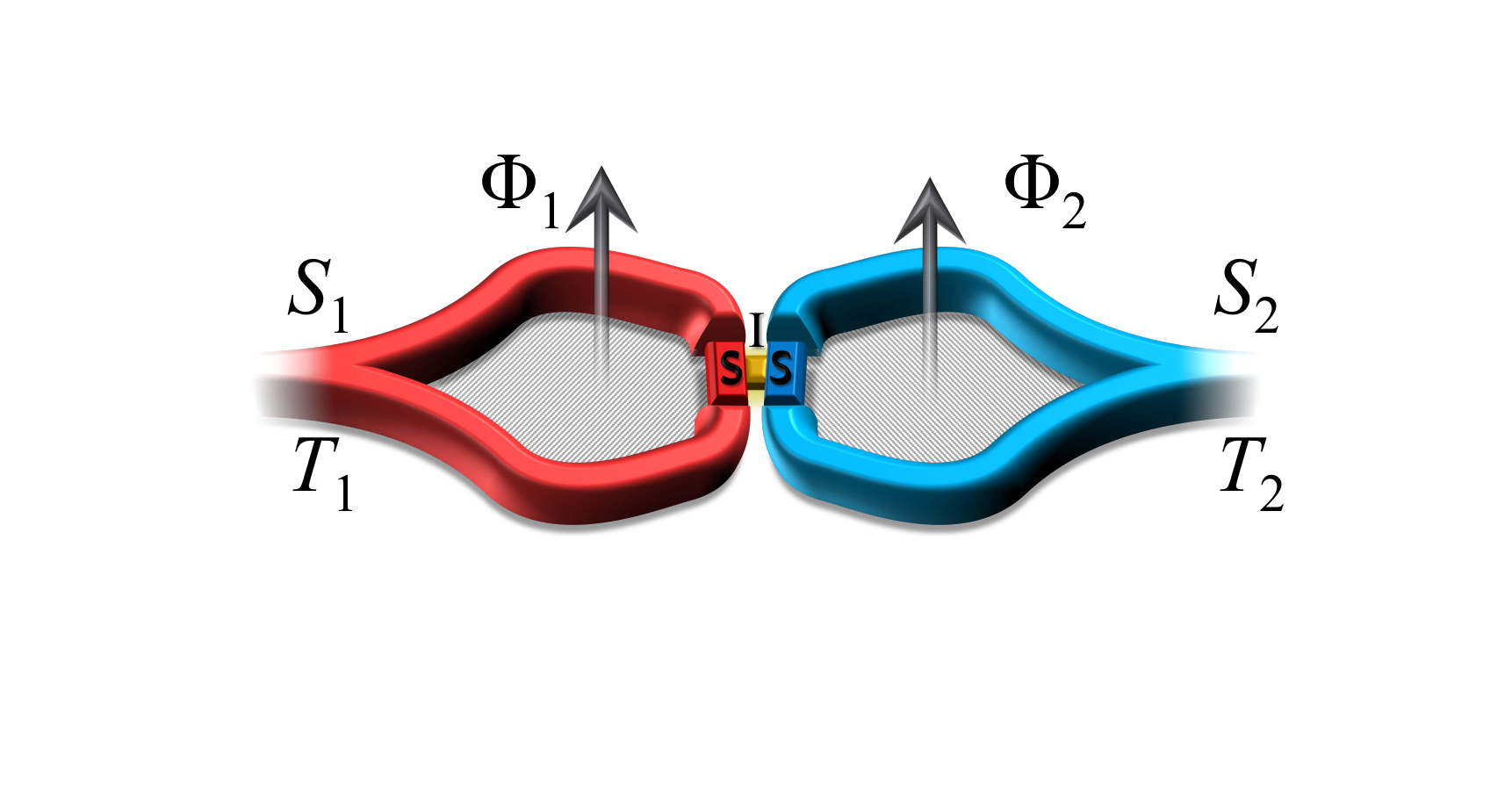}
\caption{Schematic of the device. The electrodes are formed by superconducting loops interrupted by small proximized regions, which are tunnel-coupled through an insulating barrier (yellow). The left, $S_1$, and right, $S_2$, loops reside at different temperature, $T_1$ and $T_2$, respectively, and are pierced by independent magnetic fluxes $\Phi_i ( \text{with }i=1, 2)$. }
\label{Fig01}
\end{figure}

In this work, we propose an alternative device configuration, shown in Fig.~\ref{Fig01}, formed by enclosing each electrode around a superconducting ring, and then tunnel coupling the electrodes. In this way, the magnetic flux through these loops gives an effective way to affect the superconducting gaps. 

The main aim of this work is to show the feasibility (and the benefits) of controlling the non-dissipative characteristics of a junction under thermal gradient, by using a most convenient ``knob''. Furthermore, we evince the non-trivial experimental advantage of being able to observe the phenomenology demonstrated in Ref.~\cite{GuaBra19} in a superconducting system made of only one type of material. 
In this way, it will be possible to access the wide range of parameters with just one device, instead of using many.

Tuning the device response by manipulating only the magnetic flux would be useful for verifying experimentally the reported effect, since typically the precise handling of the temperature gradient is rather demanding. In particular, we show how to magnetically control the step-like response of the critical current, and how the effect of a loop area asymmetry reflects on its tunability. The feasibility of mastering thermal transport in Josephson devices via external magnetic fluxes was already extensively demonstrated in different cases~\cite{Gia12,GuaSol18,Gua18,GuaSolBra18,Hwa18,Sen20,Bau21}. Finally, the proposed setup can operate also at a quite low temperature, thus significantly simplifying the experimental demand to verify the effect.

The paper is organized as follows. In Sec.~\ref{Sec01}, we study the behavior of the critical current by varying the magnetic fluxes driving of the device, at different values of the ratio between the critical temperatures of the two superconductors, both in the symmetric and asymmetric case. In Sec.~\ref{Sec02}, we introduce the working principles of a magnetic flux-to-critical current transducer based on a temperature-biased device, describing also a simple measurement setup. In Sec.~\ref{Sec03}, the conclusions are drawn.

\section{The critical current}
\label{Sec01}\vskip-0.2cm


The specific setup discussed in this paper is partially inspired by the superconducting quantum interference single-electron transistor (SQUISET)~\cite{Var07,Enr16,Enr17,Gre21}. The latter is composed by two identical superconducting quantum-interference proximity transistors~\cite{Gia10}, playing respectively the role of source and drain electrodes and pierced by different magnetic fluxes. Recently, a magnetometer~\cite{Fas21} and a thermal diode~\cite{Gou19} based on a SQUISET were proposed. In such a setup, the superconducting rings are closed by short normal regions, where superconducting correlations affect locally the weak-link through the proximity effect. Indeed, a phase difference in the ring induces a strong modification of the spectrum of the island, where a minigap is opened~\cite{Hei02,leS08}. The loop geometry enclosing the superconducting electrode makes it possible to change the phase difference $\varphi$ across the normal island through external magnetic fields. Thus, in the case of a short $N$ region satisfying the condition $E_{th} \gg \Delta$, where $E_{th}=\hbar D/L^2$ is the Thouless energy, with $D$ and $L$ being the diffusion constant and length of the region, respectively, both the DOS and the anomalous Green's function (AGF) can be written in a quite compact form~\cite{Bel99,Hei02,Gia11}. 

In our double-loop configuration, we envisage a tunnel barrier connecting the centers of the two proximized wires~\footnote{Following Ref.~\cite{Bel99}, we observe that in a short proximized wire in the diffusive limit the DOS and the AGF can be written in a quite compact form as $N(x,\varepsilon,T,\varphi)=\textup{Re}\left\{\text{cosh}[\theta(x,\varepsilon,T,\varphi)]\right\}$ and $\mathfrak{F}(x,\varepsilon,T,\varphi)=\textup{sinh}\left[\theta\left(x,\varepsilon,T,\varphi\right)\right]\textup{exp}\left[i\chi\left(x,\varepsilon,T,\varphi\right)\right]$, respectively. In the center of the wire ($x=0$) one obtains $\chi=0$ and $\theta(\varepsilon,T,\varphi)=\textup{arcosh} \sqrt{\varepsilon^2/\left [ \varepsilon ^2-\Delta^2 \cos^2\left ( \varphi/2 \right )\right ]}$ so that the DOS and the AGF reduce respectively to $$N(\varepsilon,T,\varphi)=\textup{Re}\left\{\varepsilon\big/\sqrt{\left [ \varepsilon ^2-\Delta^2 \cos^2\left ( \varphi/2 \right )\right ]}\right \}$$ and $$\mathfrak{F}(\varepsilon,T,\varphi)=\Delta \left| \cos\left ( \varphi/2 \right )\right|\big/\sqrt{\left [ \varepsilon ^2-\Delta^2 \cos^2\left ( \varphi/2 \right )\right ]}.$$In the case of a wire enclosed in a superconducting loop with a negligible inductance and threated by a magnetic flux $\Phi$, the phase difference becomes $\varphi=2\pi\Phi/\Phi_0$. Finally, we observe that going beyond the $x\to0$ limit is certainly possible, but this requires the introduction of other parameters that, for the sake of simplicity, we prefer not to include in this simplified discussion.}, so that we can simply assume a BCS-like spectrum with a flux-dependent gap like
\begin{equation}\label{delta_phi}
\bm{\mathit{\Updelta}}_{\Phi}(T_j,\Phi_j)=\Delta_{j}(T_j)\left | \cos \left ( \pi\Phi_j\right ) \right |.
\end{equation}
This means that the spectral properties of the two electrodes of the tunnel JJ can be tuned via the external magnetic fluxes $\Phi_j$ through the ring areas. 

A tunnel junction formed by different BCS superconductors $S_1$ and $S_2$, with energy gaps $\Delta_1$ and $\Delta_2$ and residing at temperatures $T_1$ and $T_2$, can support a Josephson current~\cite{Jos62,Jos74} with a maximum value given by the relation~\cite{Gol04,Gia05}
\begin{eqnarray}\label{IcT1T2Im}\nonumber
I_c=\frac{1}{2eR}\Bigg |\int_{-\infty}^{\infty}&& \Big \{ f_1\left ( \varepsilon \right )\textup{Re}\left [\mathfrak{F}_1(\varepsilon ) \right ]\textup{Im}\left [\mathfrak{F}_2(\varepsilon ) \right ] \\&&
+ f_2\left ( \varepsilon \right )\textup{Re}\left [\mathfrak{F}_2(\varepsilon ) \right ]\textup{Im}\left [\mathfrak{F}_1(\varepsilon ) \right ]\Big \} d\varepsilon\Bigg |.
\end{eqnarray}
Here, $R$ is the normal-state resistance of the junction and $f_j\left ( \varepsilon \right )=\tanh\left ( \varepsilon/2 k_B T_j \right )$. In the center of a proximized flux-controlled electrode, the AGF reduces to~\cite{Bar82}
\begin{equation}\label{Green}
\mathfrak{F}_j(\varepsilon) =\frac{\bm{\mathit{\Updelta}}_{\Phi}(T_j,\Phi_j)}{\sqrt{\left ( \varepsilon +i\Gamma_j \right )^2-\bm{\mathit{\Updelta}}_{\Phi}(T_j,\Phi_j)}}
\end{equation}
We included the phenomenological Dynes parameter $\Gamma_j=\gamma_j\Delta_{j0}$~\cite{Dyn78}, with $\Delta_{j0}=1.764 k_BT_{c_j}$ being the zero-temperature superconducting BCS gap~\cite{Tin04} and $T_{c_j}$ is the critical temperature. The Dynes parameter~\cite{Dyn78,Dyn84} would effectively describe a lifetime broadening, and it phenomenologically reproduces the smearing of the I-V characteristics at low voltages~\footnote{The Dynes parameter is added by introducing imaginary time component through the substitution $\varepsilon\to\varepsilon+i\Gamma$.}. In fact, a non-negligible $\gamma_j$ introduces states within the gap, $\left | \varepsilon \right |<\Delta_j$, in contrast with the ideal BCS case~\cite{Pek10,Sai12}. In this work we assume quite good quality junctions, where $\gamma_1 =\gamma_2 =\gamma \in[10^{-3}-10^{-4}]$~\cite{Pek04,Mar15,For16}.

According to Eq.~\eqref{IcT1T2Im}, the critical current of a temperature-biased tunnel JJ depends strongly on the superconductors forming the device, so that it is useful to define a gap-asymmetry parameter 
\begin{equation}\label{gap_asymmetry_parameter}
\delta=\frac{\Delta_{10}}{\Delta_{20}}=\frac{T_{c_1}}{T_{c_2}}.
\end{equation}
A suitable gap-asymmetry has been achieved by using electrodes formed by a proximity-coupled superconductor-normal metal bilayer~\cite{Ger22}, in order to fine tune the critical temperature of the film, for example by appropriately adjusting the film thickness.
Unfortunately, in such a case, to change the gap asymmetry one needs to grow another sample, so fine-tuning through the magnetic fluxes discussed below is certainly experimentally very advantageous.

For the sake of clarity, we use in this work a notation in which a tilde over a letter labels a dimensionless, normalized quantity. In particular, the quantities $\Phij=\Phi_j/\Phi_0$ (with $\Phi_0=h/2e$ being the magnetic flux quantum) and $\Tj=T_j\big/\sqrt{T_{c_1}T_{c_2}}$ are the normalized magnetic flux and the normalized temperature of the $j$-th ring, respectively, and $\Inorm=\frac{2eR}{\sqrt{\Delta_{10}\Delta_{20}}}I_c=2e\frac{\sqrt{\delta}R}{\Delta_{10}}I_c$ is the normalized critical current of the device
\footnote{The normalization with respect to the geometric mean of the critical temperatures, $\sqrt{T_{c_1}T_{c_2}}$, is convenient to treat asymmetric systems. Thus, the condition $T_1=T_2$, \emph{i.e.}, the absence of a temperature gradient, corresponds to $\Tl=\Tr$, regardless of the value of $\delta$. Anyway, in the case of asymmetric gaps, there are some constraints on the admissible normalized temperatures to keep both electrodes superconducting, that is $\Tl\leq\sqrt{\delta}$ and $ \Tr\leq 1\big/\sqrt{\delta}$}.

\begin{figure*}[t!!]
\centering
\includegraphics[width=2\columnwidth]{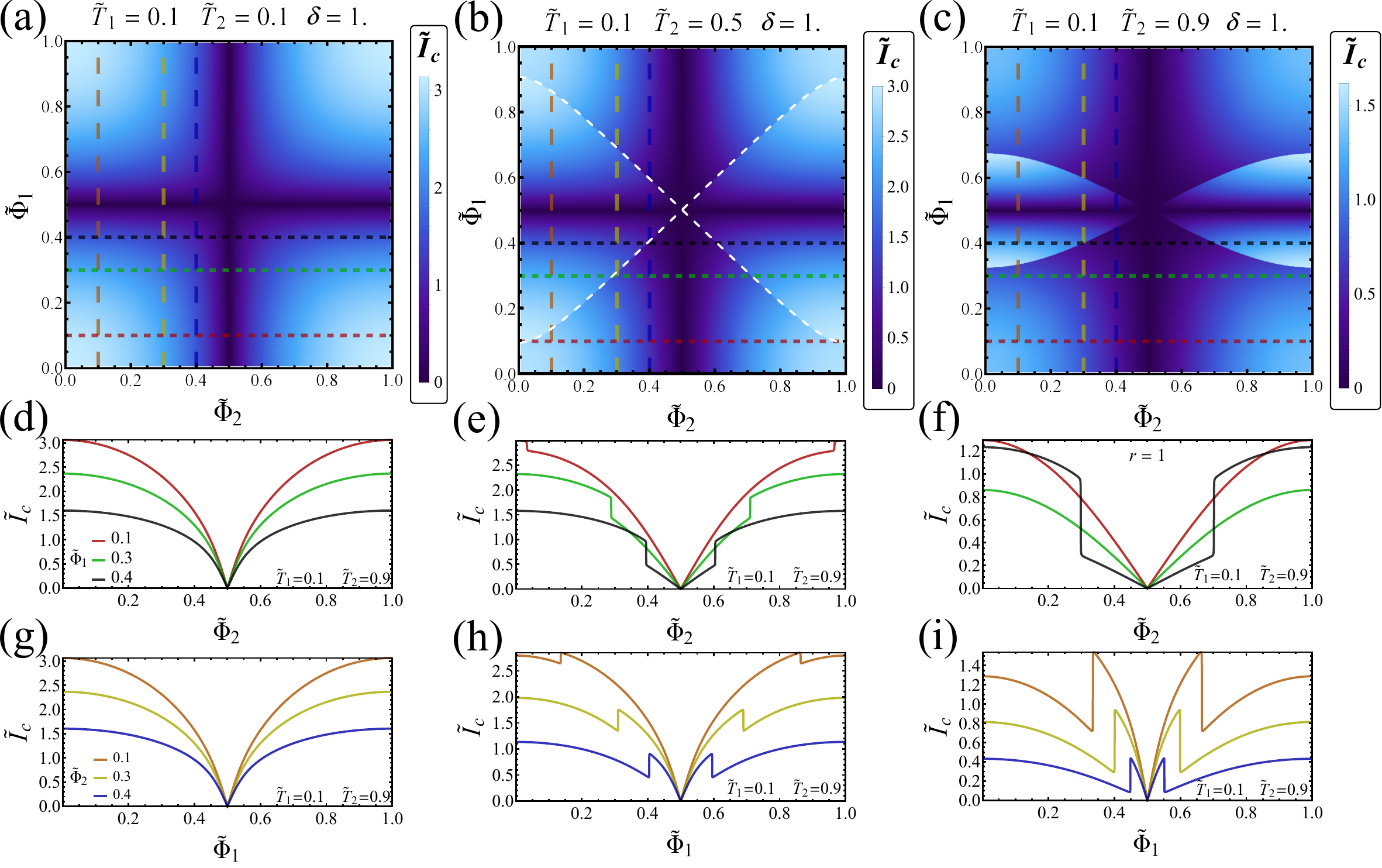}
\caption{Normalized critical current as a function of the normalized magnetic fluxes $\Phil$ and $\Phir$, at $\delta=1$, $\Tl=0.1$, $\gamma=10^{-3}$, and $\Tr=0.1, 0.5$ and $0.9$, see panels (a), (b), and (c), respectively. The horizontal short-dashed lines indicate the $\Phil$ values at which the curves in panels (d), (e), and (f) are calculated, while the vertical long-dashed lines indicate the $\Phir$ values at which the curves in panels (g), (h), and (i) are calculated. The legend in panel (d) refers also to panels (e) and (f), while the legend in panel (g) refers also to panels (h) and (i). }
\label{Fig02}
\end{figure*}

As already shown in Ref.~\cite{GuaBra19}, a temperature gradient across a JJ formed by different superconductors, \emph{i.e.}, with $\delta\neq1$, may affect its critical current in such a way to induce a steeper response. In particular, a critical current ``jump'' occurs when the electrodes reside at the temperatures at which the BCS gaps coincide and the singularities of the AGFs in the two superconductors match~\cite{Bar82}. This phenomenon is the non-dissipative counterpart of the peaks observed in the quasiparticle charge and heat current flowing through a voltage-biased $\text{S}_1$-$\text{I}$-$\text{S}_2$ junction~\cite{Sha62,Tow62,Har74,Bar82,Gol13,For16}, both determined by the alignment of the gap singularities of the BCS superconductors~\cite{Bar82}. In Ref.~\cite{GuaBra19} it was demonstrated that a requirement to observe this peculiar $I_c$ behavior is maintaining a temperature bias across the device. In other words, in these researches the fine control of temperatures is strictly necessary to observe the phenomenon. Instead, in our setup this quite demanding requirement can be relaxed through the alternative proposed design, which allows to accurately tune the superconducting gap separately. In fact, in the place of temperatures, the magnetic fluxes can be used to control the matching of singularities in the AGF, which gives rise to the critical current jumps. Moreover, the magnetic control of superconducting gaps makes it possible to observe the step-like response of $I_c$, always provided that a temperature gradient is present, even for identical zero-temperature gaps, i.e., $\delta=1$, in other words, in a device formed by electrodes of the same material.

So, in short, we point out that the crucial difference between our device and a SQUISET is that the two proximized regions are tunnel-coupled through an insulating barrier, see Fig.~\ref{Fig01}, instead to be coupled to the Coulomb-blockaded island. For our setup, the critical current of the device is still given by Eq.~\eqref{IcT1T2Im}, but the superconducting gaps depend on the magnetic fluxes according to Eq.~\eqref{delta_phi}. Consequently, the dependence on the magnetic flux of the gaps is reflected first on the AGFs $\mathfrak{F}_j$, see Eq.~\eqref{Green}, and then on the critical current $I_c$. 
Assuming the independent control of the magnetic fluxes~\cite{Enr19} or alternatively considering two different loop areas, we can investigate the system response as a function of the external magnetic flux. 
In this way, it possible to observe the anomalous $I_c$ response also when the rings are made by the same superconductor, as discussed in the following section.

\subsection{The symmetric-gap case ($\delta=1$)}
\label{Sec01a}\vskip-0.2cm

In this section we assume that the rings are made by the same superconductor, i.e., $\delta=1$. As we said, this condition cannot be considered in previous works~\cite{Gua19,GuaBra19}. Figure~\ref{Fig01} shows a possible experimental realization of the proposed device, where we indicate the magnetic fluxes, $\Phi_1$ and $\Phi_2$, piercing the superconducting rings $S_1$ and $S_2$, which reside at temperatures $T_1$ and $T_2$, respectively.

\begin{figure}[t!!]
\centering
\includegraphics[width=0.8\columnwidth]{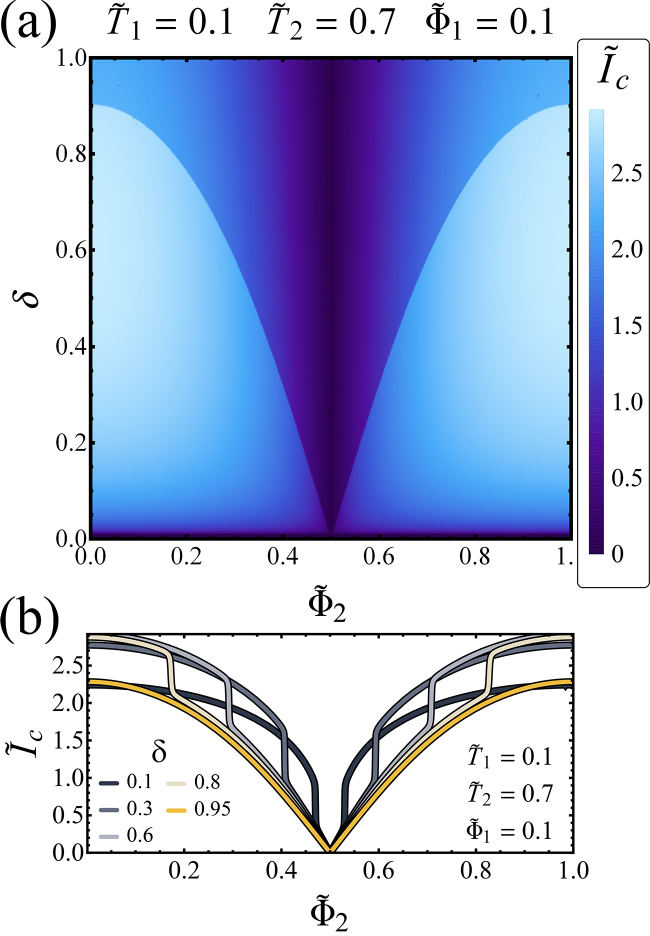}
\caption{(a) Normalized critical current as a function of $\delta$ and $\Phir$. (b) Normalized critical current as a function of $\Phir$ at different values of $\delta$. Both panels are obtained for $\Phil=0.1$, $\Tl=0.1$, $\Tr=0.7$, and $\gamma=10^{-3}$.}
\label{Fig03}
\end{figure}

The density plots in Fig.~\ref{Fig02} describe the behavior of the normalized critical current, $\Inorm$, as a function of the normalized magnetic fluxes, $\Phil$ and $\Phir$, considering normalized temperatures equal to $\Tl=0.1$ and $\Tr=0.1, 0.5,\text{ and }0.9$, see panels (a), (b), and (c), respectively~\footnote{We note that for $\delta=1$ both temperatures are normalized to the same critical temperature, i.e, $\Tj=T_j/T_c$ with $j=1,2$.}.
In all these density plots we observe that when $\Phij=0.5$ the critical current is totally suppressed, since in this case $\bm{\mathit{\Updelta}}_{\Phi}\to0$, see Eq.~\eqref{delta_phi}~\footnote{This is a clear consequence of the fact that, in general, according to the Ambegaokar-Baratoff relation~\cite{Amb63} the critical current is proportional to $I_c\propto\sqrt{\Delta_{10}\Delta_{20}}$.}. 

For comparison, we report first the results obtained assuming no thermal gradient across the system, \emph{i.e.}, $\Tl=\Tr=0.1$, see Fig.~\ref{Fig02}(a). In this situation, a magnetic flux change produces only a modulation of $\Inorm$ due to the gap closing, without the appearance of any jumps [see also Figs.~\ref{Fig02}(d) and (g)].

Conversely, the density plot in Fig.~\ref{Fig02}(b), which is obtained by increasing $\Tr$, shows an abrupt change in color, i.e., a transition from light to dark shades of blue, which indicates a sudden change of the critical current. 
This phenomenon is highlighted in panel Fig.~\ref{Fig02}(e), where three selected profiles of $\Inorm$ as a function of $\Phir$ for different $\Phil$'s are shown as well. The situations plotted in this figure corresponds to the horizontal short-dashed lines in Fig.~\ref{Fig02}(b). All curves show a step-like response of $\Inorm$ at specific values of the magnetic flux $\Phir=\Phir^*$. 
In panel (h) we show three other selected profiles of $\Inorm$ as a function of $\Phil$ for fixed $\Phir$ values, which correspond to the vertical long-dashed lines in Fig.~\ref{Fig02}(b). In this case we observe a critical current jump, but these curves behave differently with respect that in Fig.~\ref{Fig02}(e). Indeed, the profiles in Fig.~\ref{Fig02}(e) and (h), i.e., $\Inorm$ \emph{vs} $\Phir$ and $\Inorm$ \emph{vs} $\Phil$, respectively, demonstrate that $\Inorm(\Phir^*)$ reduces at the jump, while $\Inorm(\Phil^*)$ increases at the jump. 

\begin{figure*}[t!!]
\centering
\includegraphics[width=2\columnwidth]{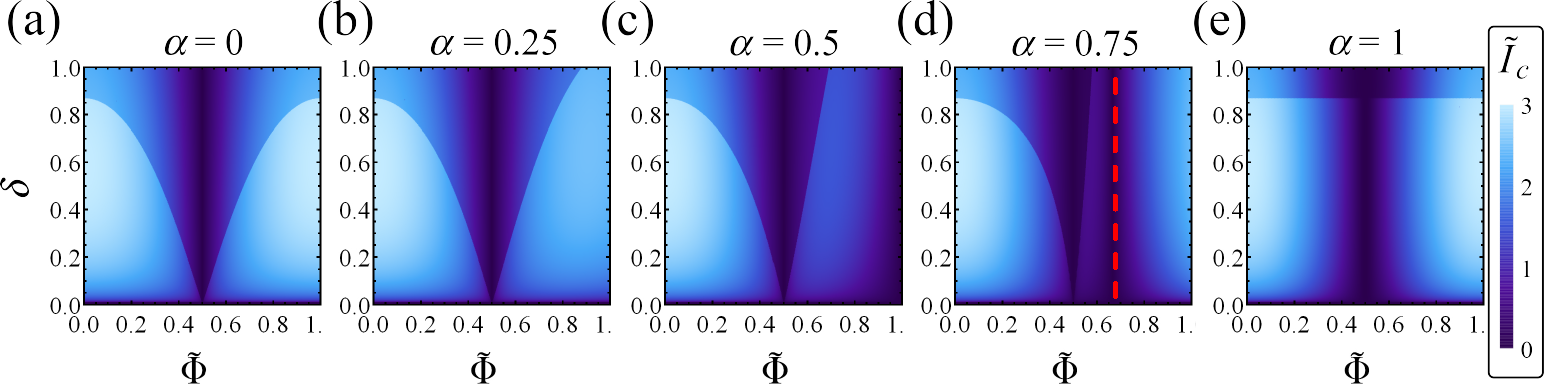}
\caption{Normalized critical current as a function of $\delta$ and $\Phi$ in the case of asymmetric area loops, with the ratio between the loops areas equal to $\alpha=\{0, 0.25, 0.5,0.75,1\}$. The values of the other parameters are: $\Tl=0.1$, $\Tr=0.7$, and $\gamma=10^{-3}$.}
\label{Fig04}
\end{figure*}

From the density plot in panels (b) and (c) we also note that the critical current undergoes a jump only within specific $\Phil$ range of values. 
To understand why in some cases the step-like behavior does not emerge, we remind that this phenomenon stems from the alignment of the singularities in the AGFs $\mathfrak{F}_j$. In other words, making explicit the magnetic flux dependence, a jump appears only when
\begin{equation}\label{gaprule}
\Delta_{1}(\Tl)\left | \cos\left ( \pi\Phil\right ) \right |=\Delta_{2}(\Tr)\left | \cos\left ( \pi\Phir\right ) \right |.
\end{equation}
In Fig.~\ref{Fig02}(b) white dashed lines mark the $(\Phil^*,\Phir^*)$ couples of values solving this equation; these values, i.e., the ``positions'' of the jumps, depend on both the working temperatures and the gap asymmetry. Conversely, for those fluxes for which Eq.~\eqref{gaprule} cannot be satisfied, the critical current shows no jumps.

\subsection{The asymmetric-gap case ($\delta\neq1$)}
\label{Sec01b}\vskip-0.2cm

Is is intriguing to discuss also how the gap asymmetry $\delta$ affects the behavior of the critical current, i.e., assuming that the electrodes are made by different superconductors, still keeping a temperature gradient between them. This is illustrated in Fig.~\ref{Fig03}(a), for $\Tl=0.1$ and $\Tr=0.7$. Here, we show how $\Inorm$ depends on $\delta$ and $\Phir$, when $\Phil$ is kept fixed, \emph{e.g.}, $\Phil=0.1$. 
We note first that for this $\Phil$ value in the symmetric-gap case ($\delta=1$) $\Inorm$ presents no jumps, see Fig.~\ref{Fig02}.
Instead, reducing the asymmetry, we can find matched cases that solve Eq.~\eqref{gaprule}: in fact, below a certain $\delta$ value, i.e., $\delta\lesssim0.9$, we observe that $\Inorm(\Phir)$ undergoes two jumps. For strong asymmetries, \emph{i.e.}, when $\delta\to0$, the positions of these two $I_c$ jumps shift towards $\Phir=0.5$. This behavior is better illustrated in Fig.~\ref{Fig03}(b), were we highlight some selected $\Inorm(\Phir)$ profiles obtained at different $\delta$.

\subsection{The asymmetric-ring area case}
\label{Sec01c}\vskip-0.2cm

In the previous sections, we assumed two independent magnetic fields piercing the superconducting rings. This strategy, although offering the control upon two independent degrees of freedom, is demanding from the point of view of the device fabrication, which has to include two independent local coils. For this reason we discuss here a simpler realization, which takes advantages on the fact that the loop areas may be different. In this way, the device can be controlled by a single magnetic field threading the areas $A_j$ of both superconducting rings, as in the case investigated in Ref.~\cite{Enr17}. In this case, we can conveniently define the area-asymmetry parameter
\begin{equation}\label{area_asymmetry_parameter}
\alpha=\frac{A_1}{A_2}=\frac{\Phil}{\Phir}.
\end{equation}
Once a temperature gradient is established, the only remaining control parameter is the common magnetic field piercing the superconducting ring areas, \emph{i.e.}, $\widetilde{\Phi}=\Phil=\alpha\,\Phir$. 

Figure~\ref{Fig04} presents a collection of density plots showing the normalized critical current $\Inorm$, in the case of loops with different areas threated by the same magnetic field, as a function of the control flux $\widetilde{\Phi}$ at different values of the gap asymmetry $\delta$. The temperature bias is kept fixed and the density plots are obtained assuming different area asymmetries $\alpha$. 
Interestingly, the condition $\alpha=0$ implies a zero area $A_1$ of the ring $S_1$. In other words, this is equivalent to replace the loop geometry with a simple superconducting stripe for the cold electrode. On the opposite side, imposing $\alpha = 1$ means taking two rings with the same area. Thus, density plots in Fig.~\ref{Fig04} serve to highlight the $\Inorm(\widetilde{\Phi},\delta)$ transition from the single-loop design, see Fig.~\ref{Fig04}(a) for $\alpha=0$, to the identical-loops design, see Fig.~\ref{Fig04}(e) for $\alpha=1$. 

We observe that even in the single-loop device the critical current behaves peculiarly, see Fig.~\ref{Fig04}(a). Then, as $\alpha$ increases, the density plot becomes more and more asymmetric in $\widetilde{\Phi}$. In particular, the right side of the contour plot, \emph{i.e.}, for $\widetilde{\Phi}>0.5$, tends to evidently distort when $\alpha\neq 0$. Conversely, the left side of the contour plot, \emph{i.e.}, for $\widetilde{\Phi}<0.5$, seems to be significantly affected by $\alpha$ only when it is very close to $1$. Interestingly, for $\alpha=0.75$ a new kink appears in the $\Inorm$ profiles [see the red dashed line in Fig.~\ref{Fig04}(d)]. 
The kinks in Fig.~\ref{Fig04} occur in correspondence of that fluxes that make the $\cos(\pi\Phi)$ or $\cos(\alpha\pi\Phi)$ terms zero. Thus, for $\alpha=0.75$ these are centered in $\Phi=1/2$ and $\Phi=1/(2\alpha)\sim0.67$.

Finally, in the identical-loops design, \emph{i.e.}, $\alpha=1$, no jumps are observed changing $\widetilde{\Phi}$, see Fig.~\ref{Fig04}(e). 
However, in this figure, we observe a clear change in the $\Inorm$ behavior, which is demonstrated by the light-to-dark transition of the density plot texture, at $\delta\simeq0.87$. Even this threshold gap asymmetry can be evaluated as that value that satisfies Eq.~\eqref{gaprule}, at a given temperature gradient.

\section{Flux-to-critical current response}
\label{Sec02}\vskip-0.2cm

The physical effects described so far could promptly find an application as a high sensitivity magnetic flux-to-critical current transducer. In this section we look at the performance of the device by adjusting the system parameters. In particular, we expect the steep change in the critical current to result in a very high sensitivity for detecting small flux changes. Indeed, a tiny variation of the magnetic flux $\Phij$ induces a huge variation of the critical current. 

When the device is employed as a magnetic flux-to-current transducer, an important figure of merit is the flux-to-current transfer function, defined as the derivative of the critical current with respect to the driving magnetic flux (critical current responsivity)~\cite{Rus11,Rus12,Esp13,Gra16}
\begin{equation}
\widetilde{I}_{\Phij}=\left | \frac{\partial \Inorm}{\partial \Phij} \right |.
\end{equation}
Since $\Inorm$ and $\Phij$ are normalized quantities, in non-normalized units the transfer function is
\begin{equation}
I_{ \Phi_j}=\frac{1}{2e\Phi_0}\frac{\Delta_{10}}{\sqrt{\delta}R} \widetilde{I}_{\Phij}.
\end{equation}
In the case of a symmetric Nb-junction, $T_{c_1}=9.2\;\text{K}$ and $\delta=1$, with $R=10\;\text{k}\Omega$, one obtains $\frac{\Delta_{10}}{2e\Phi_0\sqrt{\delta}R}\simeq0.07\;\mu\text{A}/\Phi_0$, representing the unit of measure of the transfer function.
Moreover, we can also investigate the height of the critical current jump, $\Delta \Inorm$, as a function of the various system parameters~\footnote{The height of the critical current jump, $\Delta \Inorm$, is obtained estimating first the full width at half maximum, $2\phi$, of the Lorentzian function used to fit the $\left | \partial \Inorm/\partial \Phij \right |$ profile around a peak, and then as the absolute value of the difference between the $\Inorm(\Phij)$ values calculated at a distance $\pm3\phi$ from the $\Inorm$ jump position, $\Phij^*$, that is $\Delta\Inorm=\left | \Inorm\left ( \Phij^*-3\phi \right )-\Inorm\left ( \Phij^*+3\phi \right ) \right |$}.

In the following, we consider a device formed by identical superconductors, \emph{i.e.}, $\delta=1$, and the independent control of the magnetic fields. We discuss only the figures of merit in the case of a monotonic variation of $\Inorm$ at the jump, e.g., the $\Inorm$ \emph{vs} $\Phir$ curves in Fig.~\ref{Fig02}(e) 
\footnote{We observe that in the case of a non-monotonic $\tilde{I}_c$, e.g., $\tilde{I}_c$ \emph{vs} $\tilde{\Phi}_1$ curves in Fig.~\ref{Fig02}(h), the $\tilde{I}_c$ slope changes around a jump and thus we expect a multi-peaked $\tilde{I}_{\tilde{\Phi}_1}$ transfer function, even crossing zero around the peak. This is why, in view of a flux-to-current transducer, we focus only on the analysis of the single-peaked transfer function, $\tilde{I}_{\tilde{\Phi}_2}$.}.
This means to use the magnetic flux $\Phil$ only as a control knob to tune the position of the optimal working point of the device.

\begin{figure}[t!!]
\centering
\includegraphics[width=\columnwidth]{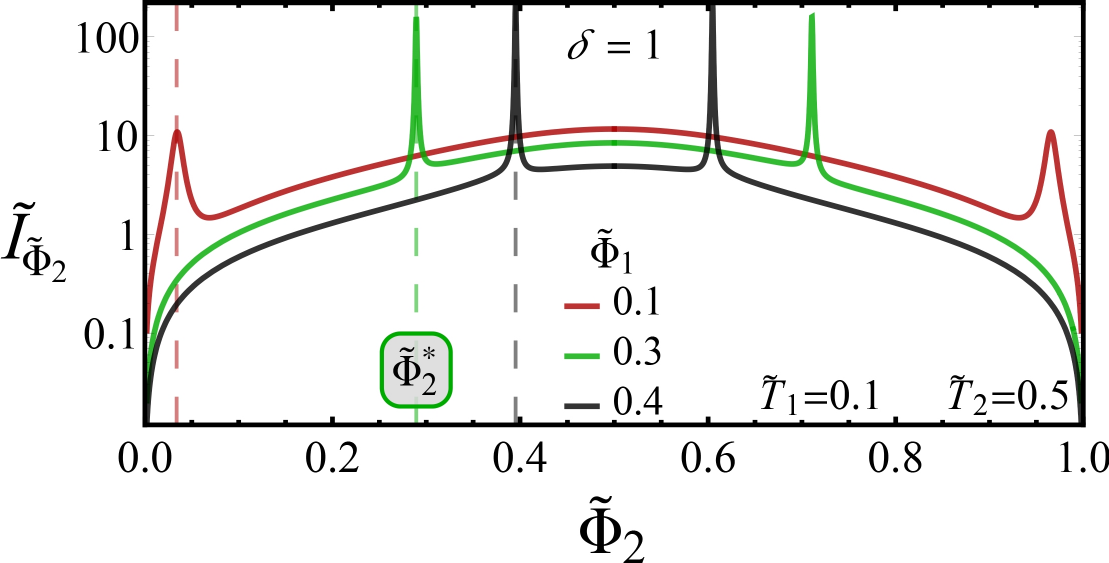}
\caption{Flux-to-current transfer functions $\widetilde{I}_{\Phir}$ at different values of $\Phil$. The values of the other parameters are: $\delta=1$, $\Tl=0.1$, $\Tr=0.5$, and $\gamma=10^{-3}$. The vertical dashed lines indicate the magnetic fluxes $\Phir^*$ at which $I_c$ jumps. The maximum transfer function, i.e., the $\widetilde{I}_{\Phir}$ value at the $\Inorm$ jump, approaches the values $\widetilde{I}_{\Phir}^{\text{max}}\simeq\{11, 160,\text{ and }220\}$ at the driving fluxes $\Phil=\{0.1, 0.3,\text{ and } 0.4\}$, respectively.}
\label{Fig05}
\end{figure}

In Fig.~\ref{Fig05} we present the $\widetilde{I}_{\Phir}$ profiles obtained by numerical differentiation of the data shown in Fig.~\ref{Fig02}(e). As expected, the transfer function is highly peaked in correspondence of each $I_c$ jump, that is when $\Phir=\Phir^*$ (these values are marked by vertical dashed lines). A magnetometer done with feedback loop kept at the optimal operating point promises a very high transfer function and, correspondingly, a very high sensitivity~\cite{Gra16}. 

\begin{figure}[t!!]
\centering
\includegraphics[width=\columnwidth]{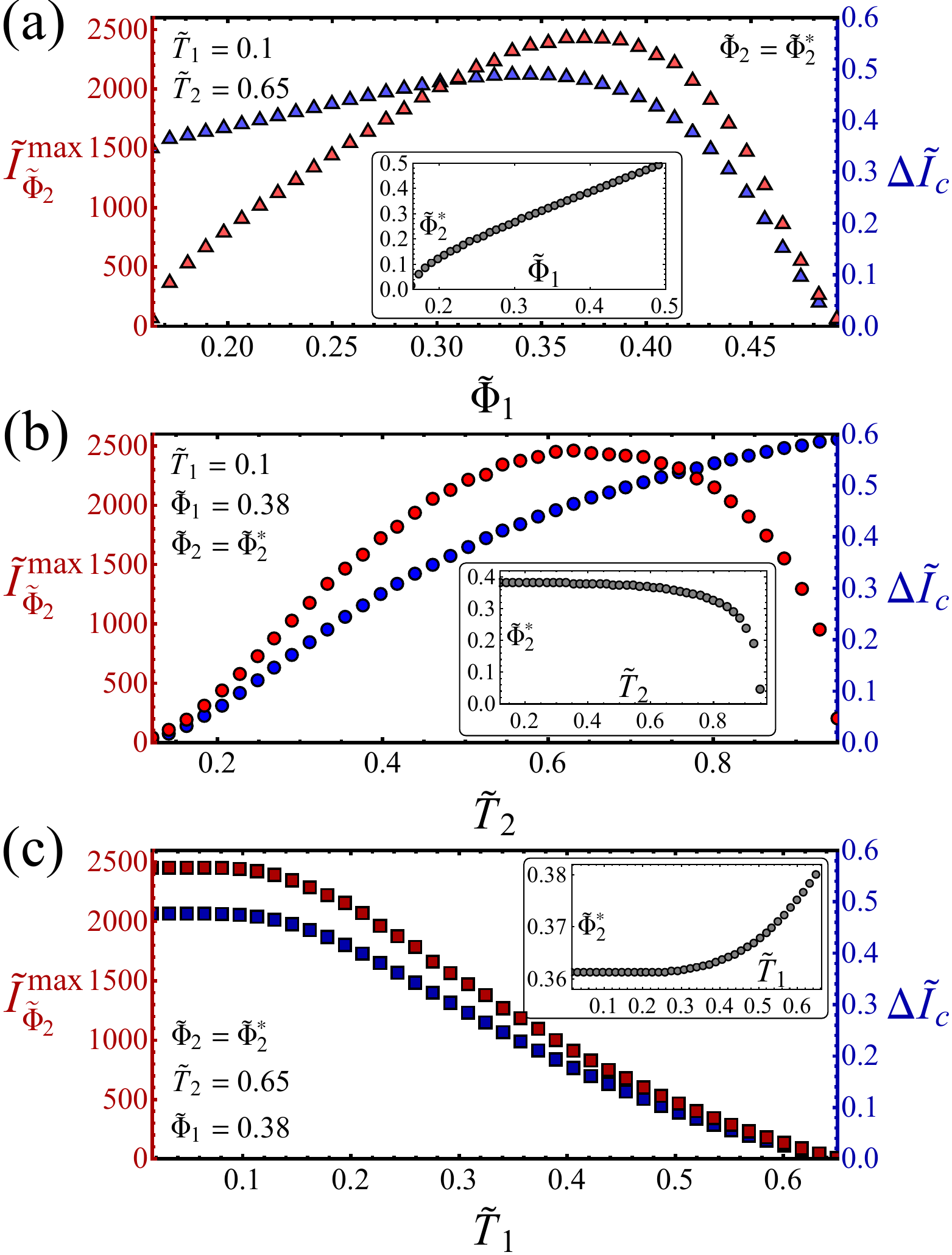}
\caption{Maximum flux-to-current transfer function $\widetilde{I}_{\Phir}^{\text{max}}$ (left axis, red symbols) and height of the critical current jump $\Delta \Inorm$ (right axis, blue symbols) versus: (a) $\Phil$ at fixed $\Tr=0.65$ and $\Tl=0.1$; (b) $\Tr$ at fixed $\Phil=0.38$ and $\Tl=0.1$; (c) $\Tl$ at fixed $\Phil=0.38$ and $\Tr=0.65$. The insets show the position, $\Phir^*$, of the $\Inorm$ jump. The other parameters are $\delta=1$ and $\gamma=10^{-4}$.}
\label{Fig07}
\end{figure}

Certainly, the height of the $\widetilde{I}_{\Phir}$ peaks depends on the steepness of the jump, which in turn depends on its ``position'', $\Phir^*$, that is on the value of the driving flux $\Phil$. In fact, from Fig.~\ref{Fig05}(a) one could figure out that when $\Phil$ tends to the value $0.5$, also $\Phir^*\to0.5$. In this case, the jump of $\Inorm$ becomes more sharp, thus making its derivative larger. However, we can reasonably expect that ${\widetilde{I}}_{\Phir}^{\text{max}}$ vanishes for $\Phil=0.5$, since at this magnetic flux the superconducting gap is suppressed. This suggests that one can search the optimal flux value as the $\Phil$ value at which the sensitivity reaches a maximum. Similarly, we can look for the optimal working temperatures, such as the temperatures that maximize the sensitivity.
Thus, we show in Fig.~\ref{Fig07}(a), (b), and (c) the behavior of both the maximum transfer function, $\widetilde{I}_{\Phir}^{\text{max}}$, (left axis, red symbols) and the height of the critical current jump, $\Delta \Inorm$, (right axis, blue symbols) as a function of $\Phil$ (at fixed $\Tr=0.65$ and $\Tl=0.1$), $\Tr$ (at fixed $\Tl=0.1$ and $\Phil=0.38$), and $\Tl$ (at fixed $\Tr=0.65$ and $\Phil=0.38$), respectively, setting $\delta=1$ and $\gamma=10^{-4}$.

\begin{figure}[t!!]
\centering
\includegraphics[width=\columnwidth]{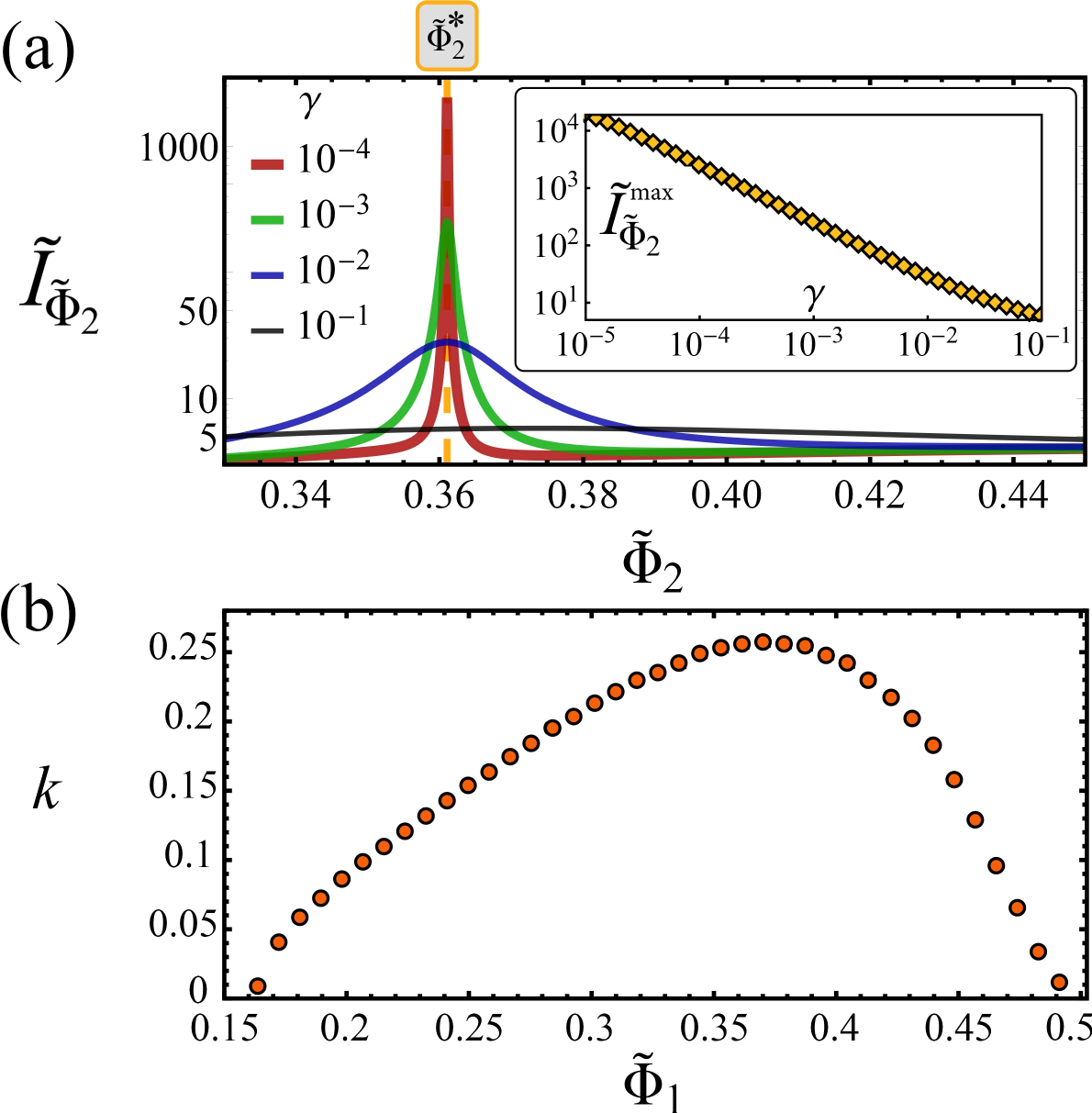}
\caption{Flux-to-current transfer function $\widetilde{I}_{\Phir}$ as a function of $\Phir$ at different values of the Dynes parameter $\gamma$, at a fixed $\Phil=0.38$. The magnetic flux $\Phir^*$ at which $\Inorm$ shows a jump is highlighted. In the inset: Maximum flux-to-current transfer functions $\widetilde{I}_{\Phir}^{\text{max}}$ (\emph{i.e.}, calculated at $\Phir=\Phir^*$) as a function of $\gamma$. (b) Fitting parameter $k$ as a function of $\Phil$, see Eq.~\eqref{fit}. The other parameters are: $\delta=1$, $\Tl=0.1$, $\Tr=0.65$.}
\label{Fig06b}
\end{figure}

Both $\widetilde{I}_{\Phir}^{\text{max}}$ and $\Delta \Inorm$ \emph{vs} $\Phil$, see Fig.~\ref{Fig07}(a), behave non-monotonically. At low $\Phil$, the $\Inorm$ jump is quite smooth being placed close to $\Phir^*\sim 0$ [see the inset of Fig.~\ref{Fig07}(a)], but despite this its height $\Delta \Inorm$ is still sizable. On the other side, for $\Phil\to0.5$ the jump position tends to $\Phir^*\to 0.5$, so that the critical current tends to vanish due to the reduced gap, and both $\widetilde{I}_{\Phir}^{\text{max}}$ and $\Delta \Inorm$ tend to zero. 

A non-monotonicity characterizes also the $\widetilde{I}_{\Phir}^{\text{max}}$ \emph{vs} $\Tr$ curve in Fig.~\ref{Fig07}(b), whereas $\Delta \Inorm$ tends to increase with $\Tr$. Clearly, both figures of merit vanish for low $\Tr$'s, since for $\Tr=\Tl$ there is no thermal gradient along the system and, thus, no critical current jump. On the other side, at large $\Tr$ values the jump position $\Phir^*$ shifts towards zero, see the inset of Fig.~\ref{Fig07}(b), thus making the $\Inorm$ step smoother, but still sizable: this is why $\Delta \Inorm$ remains quite large even at high $\Tr$ values. 

Finally, increasing $\Tl$, see Fig.~\ref{Fig07}(c), both $\widetilde{I}_{\Phir}^{\text{max}}$ and $\Delta \Inorm$ show a plateau at low $\Tl$ and then reduce until vanishing for $\Tl=\Tr$. This is clearly due the fact that lowering the thermal gradient reduces the critical current jump~\cite{GuaBra19}. We observe also that the jump position $\Phir^*$ is only weakly affected by a change of $\Tl$, as shown in the inset of Fig.~\ref{Fig07}(c).

In summary, the flux-to-current transfer function $\widetilde{I}_{\Phir}^{\text{max}}$ of a symmetric device, i.e., with $\delta=1$, can be maximized by making the electrodes to reside at the temperatures $\Tr\simeq0.65$ and $\Tl\simeq0.1$ and setting the driving magnetic flux to the value $\Phil\simeq0.38$. In this case, $\widetilde{I}_{\Phir}^{\text{max}}\sim2500$, that, in non-normalized units and assuming $T_{c_1}=9.2\;\text{K}$, $\delta=1$, $R=10\;\text{k}\Omega$, and $\gamma =10^{-4}$, is equal to $ I_{\Phi_2}^{\text{max}}\sim175\;\mu\text{A}/\Phi_0$~\footnote{We additionally stress that the sensitivity can be further optimized by assuming an asymmetric device, that is made by different superconductors. In fact, the value $\tilde{I}_{\Phir}^{\text{max}}\sim3400$, corresponding to $ I_{\Phi_2}^{\text{max}}\sim240\mu\text{A}/\Phi_0$, is reached in the case of $\delta=0.4$, assuming optimal values for the other parameters (plot not shown).}.

Since the sensitivity of the device essentially relies upon the sharpness of the $\Inorm$ jump, it in turn depends also on the value of the phenomenological Dynes's parameters~\cite{GuaBra19}. Thus, in Fig.~\ref{Fig06b}(a) we show the behavior of $\widetilde{I}_{\Phir}$ \emph{vs} $\Phir$ in a neighborhood of a jump, at a few values of $\gamma$ and $\Tl=0.1$, $\Tr=0.65$, and $\Phil=0.38$. As expected, the lower $\gamma$, the more peaked is $\widetilde{I}_{\Phir}$: in particular, as it is demonstrated clearly in the inset of Fig.~\ref{Fig06b}(a), the maximum transfer function decreases by increasing $\gamma$ as
\begin{equation}\label{fit}
\widetilde{I}_{\Phir}^{\text{max}}\simeq \frac{k}{\gamma}
\end{equation}
(see Appendix~\ref{AppA} for more details about the origin of the $1/\gamma$ dependence of the critical current responsivity).
From the fit of the data in the inset of Fig.~\ref{Fig06b}(a) we obtain the value $k\simeq0.26$. Being the fitting coefficient $k$ in Eq.~\eqref{fit} a function of the jump position, a given driving flux $\Phil$ matches a specific value of $k$, see Fig.~\ref{Fig06b}(b). Therefore, according to Eq.~\eqref{fit}, an experimental measurement of the transfer function at the matching conditions could, in principle, give an estimate of the Dynes parameter $\gamma$.

\begin{figure}[t!!]
\centering
\includegraphics[width=\columnwidth]{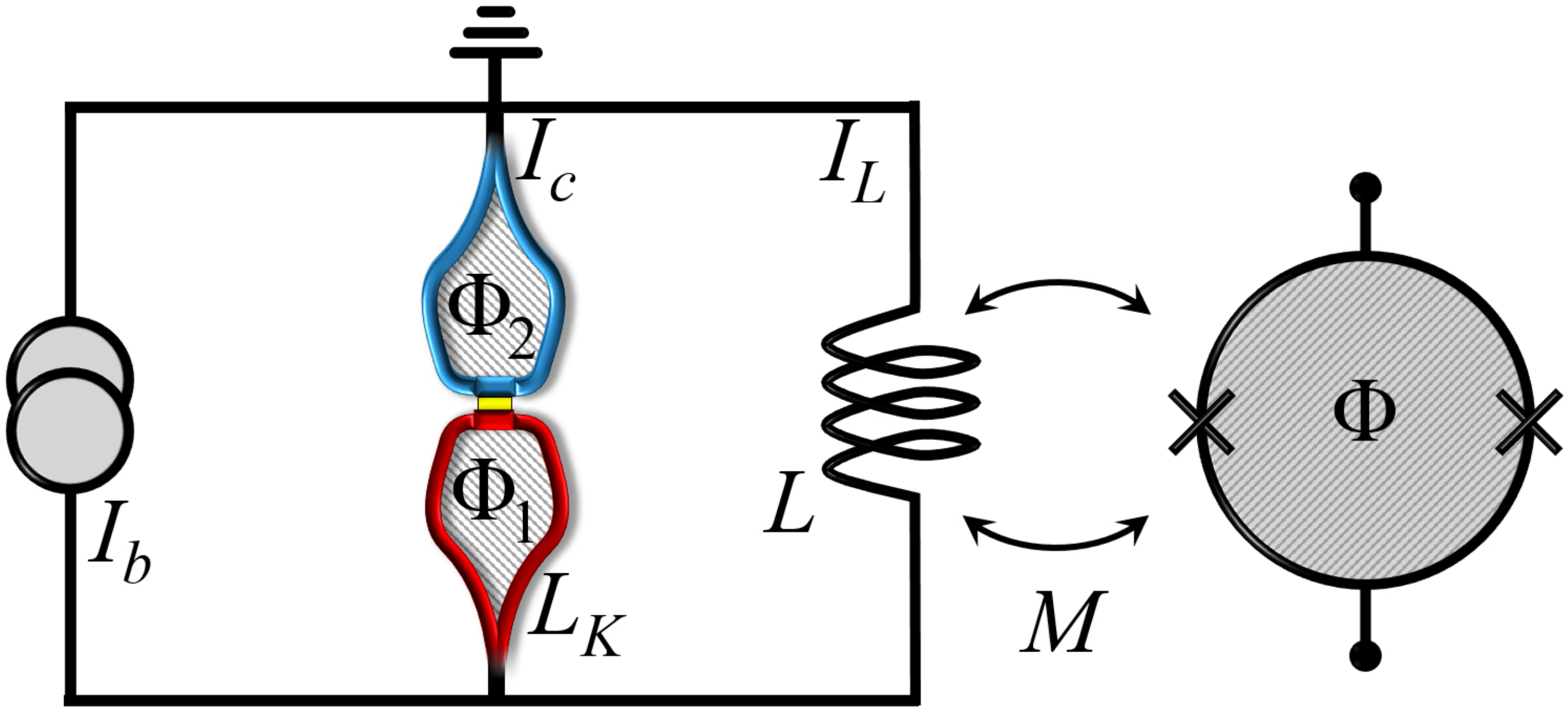}
\caption{Readout scheme including a dc SQUID.}
\label{Fig08}
\end{figure}

Finally, we observe that the information content we are interested in can be extracted through a scheme, depicted in Fig~\ref{Fig08}, that uses a standard dc-SQUID, drawn in gray in the figure, along the lines of the readout strategy for proximity Josephson sensors discussed in Ref.~\cite{Gia08}. In such a scheme the current flowing in the inductance $L$ is changed when the critical current $I_c$ in the double-ring device is modified. In fact, a constant bias current $I_b$ divides into two parts, i.e., one flowing through the double-ring branch and the other ($I_L$) through a load inductor ($L$), which is mutually coupled to the dc SQUID. The variation of the critical current $I_c$ produces a large enhancement of the Josephson kinetic inductance $L_K=\hbar/(2eI_c)$, which in turn results in a modification of $I_L$, thus producing a magnetic field which is detected by the SQUID. Other schemes, which resort to dispersive measurements, are also possible, but we do not investigate them further.

In conclusion, we observe that the flux-to-critical-current transducer discussed in this work shares the same applicative contexts of SQUIDs and phase coherent amplifiers. Here, the key figure of merit is the flux-to-current response function, which in a simple SQUID typically depends only on the critical currents and the asymmetry between the two embedded JJs; these physical parameters are fixed and given by the device-specific implementation. Instead, in our case we have demonstrated that the figure of merit can be ``controlled'' also by the temperature gradient, the Dynes parameters, or simply by fixing the operating point tuning the controlling flux. This feasibility to adjust the flux-to-current response is a crucial advantage when the dispersive detection is eventually adopted. We are confident that the control flexibility of this device may be beneficial in different context, such as the phase coherent amplification or even in the quantum sensing of temperature and/or temperature gradient. 

\section{Conclusions}
\label{Sec03}\vskip-0.2cm

In conclusion, in this paper we explored the behavior of the critical current, $I_c$, of a Josephson device with a very specific design. In fact, we considered a tunnel Josephson junction formed by proximized electrodes enclosed in two different superconducting loops pierced by independent magnetic fluxes, which allow the independent tuning of the superconducting gaps. We discussed the device response by changing both the temperatures of the electrodes and the ratio, $\delta$, between the critical temperatures of the superconductors. 
In particular, at a fixed temperature bias, we discuss the step-like response of $I_c$ when the system is driven by magnetic fluxes at which the BCS superconducting gaps coincide. Markedly, unlike the asymmetrical junction case discussed in Ref.~\cite{GuaBra19}, this peculiar step-like response appears even in a symmetric device, that is formed by electrodes of the same material, i.e., $\delta=1$.

We also discussed the optimal working conditions to increase the transfer function of a magnetic flux-to-critical current transducer, considering even the relevant figures of merit. In this case, we also illustrated how the value of the Dynes parameter in the superconductors influences the sharpness of the $I_c$ transition, being an important factor to increase the sensitivity.

The proposed setup can be imaged as a novel superconducting flux transducer~\cite{Gra16}, where the sensitivity to magnetic fluxes is determined by non-equilibrium temperature conditions. The feasibility of controlling the superconducting gaps, and therefore the transport properties, by means of magnetic fluxes allows to relax the rather tight requirement of fine tuning of the temperatures of the two electrodes previously discussed~\cite{GuaBra19,Gua19}.
Indeed, once a temperature difference across the system has been established, we only need to adjust the magnetic fluxes to properly establish the operating point, so to highlight the steep jump of the critical current. Furthermore, we can search for the best conditions that give a more abrupt response, or a more intense jump, simply by changing the magnetic flux, thus making the proposed device flexibility very appealing.

Finally, although the showed peculiar response of the critical current has only been discussed theoretically and it was not yet observed experimentally, we should note that under similar (not identical) conditions (i.e., an asymmetric temperature biased junction) a thermoelectric effect on the quasiparticle current~\cite{MarPRL20} has been recently observed experimentally~\cite{Ger22}. However, in this case the focus is on the dissipative current, whereas the Josephson current must be suppressed. So, even if there is still not a ``fine'' temperature control, in the proposed setup we think the line is traced and we hope that the Josephson current can be soon investigated. 
Our proposal offers a possible experimental route to verify the effect without the requirement of highly detailed control of the junction electrode temperatures.

\appendix

\section{Scaling of the critical current responsivity, the role of $\gamma$}
\label{AppA}

In this section we delve into the role of the Dynes parameter, with a focus on the origin of the scaling of the transfer function Eq.~\eqref{fit} with $\gamma$ around the matching point. 

Actually, in our work we have chosen $\gamma$ values lying in a range often used for good quality superconducting systems; moreover, these values seem to predict well the behavior of quasiparticle current in realistic device. 
Despite its phenomenological nature, the Dynes-like picture has been shown to give an excellent match for experimentally observed DOS in a wide variety of diverse superconducting systems. 
However, at the moment there is not enough information on its value, at least from the behavior of the non-dissipative current.
Thus, the measurement of the device proposed in this work could potentially shed light also on the role of this phenomenological parameter in the non-dissipative regime.

In the inset of Fig.~7(a), we observed that the derivative of the critical current with respect to the driving magnetic flux, $\left | \frac{\partial \Inorm}{\partial \Phij} \right |$, when the other magnetic flux ($\Phii$) is kept fixed, scales as $1/\gamma$ when the superconducting gaps coincide, see Eq.\eqref{delta_phi}. Firstly, we note that the critical current can be written as the sum of two terms, i.e., $I_c=I_{ij}+I_{ji}$, with $i,j=1,2$, where
\begin{eqnarray}\label{AppAEq0}
I_{ij}=\frac{1}{2eR}\int_{-\infty}^{\infty}f_i\left ( \varepsilon \right )\mathfrak{G}_{ij}(\varepsilon,\Delta_i,\Delta_j)d\varepsilon,
\end{eqnarray}
$f_i\left ( \varepsilon \right )=\tanh\left ( \varepsilon/2 k_B T_i \right )$, and we have defined the function $\mathfrak{G}_{ij}(\varepsilon,\Delta_i,\Delta_j)=\textup{Re}\left [\mathfrak{F}_i (\varepsilon,\Delta_i)\right ]\textup{Im}\left [\mathfrak{F}_j(\varepsilon,\Delta_j) \right ]$ in terms of the AGFs and stressing the role of the superconducting gaps.

\begin{figure}[t!!]
\centering
\includegraphics[width=0.85\columnwidth]{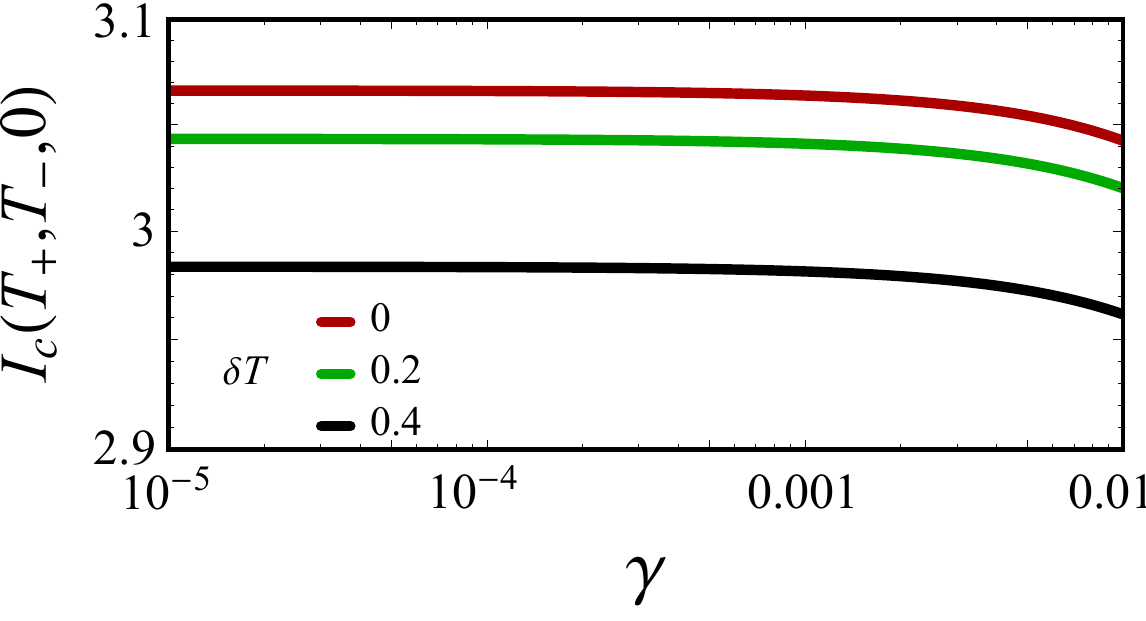}
\caption{$I_c(T_{+},T_{-},0)$ versus $\gamma$, with $T_{\pm}=T\pm\delta T/2$, at $T=0.4$ and $\delta T=\{0, 0.2,0.4\}$.}
\label{Fig01App}
\end{figure}

At the matching point, $\Delta_i=\Delta_j=\Delta$, it is convenient to express the energies in terms of $\Delta$, such as $\epsilon=\varepsilon/\Delta$, and the current in unit of $\Delta/(2eR)$. In this case, the product between real and imaginary parts of the AGFs reduces to a simple combination of Lorentzian functions:
\begin{equation}\label{AppAEq1}
\mathfrak{G}_{0}(\epsilon)= \frac{\gamma}{4}\left [ \frac{1}{\gamma ^2+(\epsilon-1)^2}-\frac{1}{\gamma ^2+(\epsilon+1)^2} \right ].
\end{equation}
This simple expression holds only for a specific case, i.e., at the matching point, so it tells us nothing about how the critical current behaves when the gaps slighlty differ, for example due to a magnetic flux modulation. To better explore this situation, it is convenient to expand $\mathfrak{G}_{ij}$ and $\mathfrak{G}_{ji}$ around the matching-gap condition, by setting $\Delta_i=\Delta$ and expanding $\Delta_j=\Delta(1+\mathcal{D})$. In this way, in normalized units, one can write
\begin{eqnarray}\label{AppAEq2}
\mathfrak{G}_{ij}(\epsilon,\mathcal{D})&\simeq& \mathfrak{G}_0(\epsilon)+\partial_{{\scriptscriptstyle \mathcal{D}}} \mathfrak{G}_{+}\;\mathcal{D}\\
\mathfrak{G}_{ji}(\epsilon,\mathcal{D})&\simeq& \mathfrak{G}_0(\epsilon)+\partial_{{\scriptscriptstyle \mathcal{D}}} \mathfrak{G}_{-}\;\mathcal{D}
\end{eqnarray}
where
\begin{eqnarray}\label{AppAEq3}
\partial_{{\scriptscriptstyle \mathcal{D}}} \mathfrak{G}_{\pm}=&&\frac{i}{4}\! \left(\frac{1}{\sqrt{(\epsilon +i \gamma )^2-1}}\pm\frac{1}{\sqrt{(\epsilon -i \gamma )^2-1}}\right)\\\nonumber
&&\!\times\left(\frac{(\epsilon+i\gamma )^2}{\left[(\epsilon+i\gamma )^2-1\right]^{3/2}}\mp\frac{(\epsilon-i\gamma )^2}{\left[(\epsilon-i\gamma )^2-1\right]^{3/2}}\right)\!.
\end{eqnarray}
Indicating, for the sake of convenience, the temperatures $(T_i,T_j)$ as $(T_{+},T_{-})$, the critical current can be approximated as 
\begin{eqnarray}\label{AppAEq4}
I_c\left ( T_{+},T_{-},\mathcal{D}\right )\!\simeq\! I_{c,0}(T_{+},T_{-})\!+\!\mathcal{D}\!\sum_{s=\pm}\!\delta I_{s}(T_{s})\!+\!\mathcal{O}(\mathcal{D}^2)\!,\qquad
\end{eqnarray}
where 
\begin{eqnarray}\label{AppAEq5}
I_{c,0}(T_{+},T_{-})&=&\int_{-\infty}^{\infty}\mathfrak{G}_0(\epsilon) \left [ f(\epsilon , T_{+})+f(\epsilon , T_{-}) \right ]d\epsilon \qquad\quad\\
\delta I_{\pm}(T_{\pm})&=&\int_{-\infty}^{\infty}\partial_{{\scriptscriptstyle \mathcal{D}}} \mathfrak{G}_{\pm}\;f(\epsilon , T_{\pm})d\epsilon.
\end{eqnarray}
Interestingly, $I_c(T_{+},T_{-},0)$ depends very little on $\gamma$, see Fig.~\ref{Fig01App} where we set $T_{\pm}=T\pm\delta T/2$ and we vary $\delta T$. This behavior reflects the general fact that at the matching point the value of the critical current is weakly dependent on $\gamma$. See, for example, Fig.~4 of Ref.~\cite{GuaBra19}, where one gap deviates from the matching condition due to the temperature change of a superconducting electrode. This happens also when the variation of the gap is due to a change in the magnetic flux, see Eq.~\eqref{delta_phi}, as considered in the present case.

In the low-temperature limit, $T\ll\Delta$, the $\tanh\left ( \epsilon/ T \right )$ can be approximated with the $\text{sgn}\left( \epsilon \right )$ function, so that the integrals $\delta I_{\pm}(T_{\pm})$ can be even solved analytically.
Anyway, in the absence of a temperature gradient, i.e, $T_{+}=T_{-}=T$, one finds that $\delta I_{+}(T)+\delta I_{-}(T)\simeq\pi/2$. This means that the change of the critical current from the matching condition value is independent from the Dynes parameter, at least in the range of $\gamma$ values physically relevant and considered in the manuscript. 

In order to observe the steep-like behavior of $I_c$, a temperature gradient across the system is necessary, i.e., $T_{+}\neq T_{-}$. So, it is interesting to investigate if the presence of a temperature gradient can eventually determine also the reported $\gamma$-scaling. Thus, to retrieve it, we can first look separately at the behavior of the functions $\delta I_{\pm}(T_{\pm})$. Using again the approximation $\tanh\left ( \epsilon/ T_{\pm} \right )\sim\text{sgn}\left( \epsilon \right )$ we can calculate the $\delta I_{\pm}$ analytically, thus allowing us to immediately appreciate how they vary with $\gamma$ 
\begin{eqnarray}\label{AppAEq6}
\delta I_{\pm}&\simeq&\frac{1}{2 \gamma }\!\left[\frac{ \pi \gamma -2 \gamma \tan ^{-1}(\gamma )}{2}+\frac{\gamma ^2\pm1}{\gamma ^2+1 }\right]=\frac{\mathcal{I}_\pm(\gamma)}{\gamma },\quad\qquad
\end{eqnarray}
where the functions $\mathcal{I}_\pm(\gamma)\simeq1/2+\mathcal{O}(\gamma)^2$ are almost constant for small $\gamma$ values (that is, in non-normalized units, for a Dynes parameter much smaller than the superconducting gaps). Finally, when adding the $\delta I_\pm$ contributions into Eq.~\eqref{AppAEq4}, in the presence of a temperature gradient the $1/\gamma$ scaling in the critical current survives, as discussed in the main text. To conclude, we stress that Eq.~\eqref{AppAEq6} has been derived assuming a deviation from the matching condition, regardless of what may cause this modulation. Thus, the result obtained is general, being valid whether we consider the effect of temperature~\cite{GuaBra19}, magnetic flux changes, or any other mechanism that could affect the superconducting gap. 

\begin{acknowledgments}
F.G. acknowledges the European Research Council under Grant Agreement No. 899315-TERASEC, and the EU’s Horizon 2020 research and innovation program under Grant Agreement No. 800923 (SUPERTED) and No. 964398 (SUPERGATE) for partial financial support. 
A.B. acknowledges the SNS-WIS joint lab QUANTRA, funded by the Italian Ministry of Foreign Affairs and International Cooperation and the Royal Society through the International Exchanges between the UK and Italy (Grants No. IEC R2 192166 and IEC R2 212041).
\end{acknowledgments}


%

\end{document}